\documentclass[11pt]{article}
\usepackage{geometry}  
\usepackage[T1]{fontenc}    
\usepackage[utf8x]{inputenc}
\usepackage{fancyhdr}       
\usepackage{enumitem}       
\usepackage{hyperref}       
\usepackage{titling}        
\usepackage[numbers]{natbib}
\usepackage{mathtools}      
\usepackage{titlesec}       
\usepackage{lastpage}       
\usepackage[english]{babel} 
\usepackage{amsmath}        
\usepackage{amsfonts}       
\usepackage{amssymb}        
\usepackage{array}          
\usepackage{color}          
\usepackage{algorithm, algorithmic, caption}
\usepackage{xspace}         
\usepackage{booktabs}
\usepackage{multirow}
\usepackage{siunitx}        
\usepackage{textcomp}       
\usepackage{bm}             
\usepackage{graphicx}
\usepackage[normalem]{ulem}

\hypersetup{%
  pdfborderstyle={/S/U/W 1}
}

\setcitestyle{round,numbers}
\makeatletter
\def\BState{\State\hskip-\ALG@thistlm}
\makeatother

\titleformat{name=\section}{\normalfont\LARGE\bfseries}{}{0pt}{}
\titleformat{name=\subsection}{\normalfont\Large\bfseries}{}{0pt}{}
\titleformat{name=\subsubsection}{\normalfont\large\bfseries}{}{0pt}{}

\newcommand{\email}[1]{\href{mailto:{#1}}{{#1}}}

\newcommand{\keywords}[1]{\textbf{Keywords}: {#1}}

\newcommand{\optinput}[1]{}

\newtagform{brackets}{[}{]}
\usetagform{brackets}

\newcommand{\thejournal}[1]{Magnetic Resonance in Medicine}

\newcommand{\herm}{^{\prime}}
\newcommand{\frob}{_\textsf{F}}

\title{High-dimensional Fast Convolutional Framework (HICU) for Calibrationless MRI}

\begin{document}

\begin{titlepage}
	\begin{center}
	    \sf{Published in Magnetic Resonance in Medicine}\\
		\bigskip
		{\noindent\LARGE\bf \thetitle}
	\end{center}
\bigskip

\begin{center}\large
    Shen Zhao\textsuperscript{1},
    Lee C. Potter\textsuperscript{1,2}
    Rizwan Ahmad\textsuperscript{*1,2,3}
\end{center}

\bigskip

\noindent

\begin{enumerate}[label=\textbf{\arabic*}]
\item Electrical and Computer Engineering, The Ohio State University, Columbus OH, USA
\item Davis Heart \& Lung Research Institute, The Ohio State University, Columbus OH, USA
\item Biomedical Engineering, The Ohio State University, Columbus OH, USA
\end{enumerate}

\bigskip
\textbf{*} Corresponding author:

\indent\indent
\begin{tabular}{>{\bfseries}rl}
Name        & Rizwan Ahmad                                                      \\
Department  & Biomedical Engineering                                            \\
Institute   & The Ohio State University                                         \\
Address     & 460 W 12th Ave, Room 318                                          \\
            & Columbus OH 43210, USA                                            \\
E-mail      & \email{ahmad.46@osu.edu}                                          \\
\end{tabular}

\vfill
\begin{tabular}{>{\bfseries}rl}
	Manuscript Word Count:	& 4,842\\
	Abstract  Word Count:	& 170 \\
\end{tabular}

\end{titlepage}
\pagebreak

\begin{abstract}

\noindent
\textbf{Purpose:} To present a computational procedure for accelerated, calibrationless magnetic resonance image (Cl-MRI) reconstruction that is fast, memory efficient, and scales to high-dimensional imaging.\\
\textbf{Theory and Methods:} Cl-MRI methods can enable high acceleration rates and flexible sampling patterns, but their clinical application is limited by computational complexity and large memory footprint. The proposed computational procedure, HIgh-dimensional fast ConvolUtional framework (HICU), provides fast, memory-efficient recovery of unsampled k-space points. For demonstration, HICU is applied to six 2D T2-weighted brain, seven 2D cardiac cine, five 3D knee, and one multi-shot diffusion weighted imaging (MSDWI) datasets.\\ 
\textbf{Results:} The 2D imaging results show that HICU can offer one to two orders of magnitude computation speedup compared to other Cl-MRI methods without sacrificing imaging quality. The 2D cine and 3D imaging results show that the computational acceleration techniques included in HICU yield computing time on par with SENSE-based compressed sensing methods with up to 3\,dB improvement in signal-to-error ratio and better perceptual quality. The MSDWI results demonstrate the feasibility of HICU for a challenging multi-shot echo-planar imaging application.\\
\textbf{Conclusions:} The presented method, HICU, offers efficient computation and scalability as well as extendibility to a wide variety of MRI applications.
\end{abstract}

\bigskip

\bigskip
\textbf{Running title:} Fast calibrationless MRI
\pagebreak

\section{INTRODUCTION}
In magnetic resonance imaging (MRI), the k-space data are samples from the continuous Fourier transform of the underlying image. Physical and physiological limits (e.g., gradient slew rate and nerve stimulation) impede MRI acquisition efficiency, leading to long acquisition times, especially for applications that require a large corpus of k-space data such as volumetric, dynamic, or high-resolution imaging. The long acquisition time can compromise patient comfort, reduce throughput, and increase motion artifacts. Therefore, image reconstruction using less data but without losing image quality has been an active research area for three decades.

Image recovery from partially sampled k-space is typically facilitated by parallel MRI (pMRI), which is available on almost all clinical scanners. In pMRI, each of several receive coils provides k-space samples of the image modulated by a corresponding spatial sensitivity map. Sensitivity Encoding (SENSE) \cite{pruessmann1999sense} and GeneRalized Autocalibrating Partial Parallel Acquisition (GRAPPA) \cite{griswold2002generalized} are the most commonly used pMRI reconstruction methods. SENSE uses the coil sensitivity maps to solve for the image via linear least-squares; however, SENSE reconstruction assumes sensitivity maps are known, which typically requires the acquisition of a separate scan or a fully sampled auto-calibration signal (ACS) region. 
By leveraging the linear dependence of one k-space sample on its neighboring samples from all coils, GRAPPA 
employs interpolation kernels learned from the ACS region to fill the missing k-space.
Iterative Self-consistent Parallel Imaging Reconstruction from Arbitrary K-space (SPIRiT) \cite{lustig2010spirit}, a generalization of GRAPPA, reorganizes interpolators into annihilating kernels.
Parallel Reconstruction Using Null Operations (PRUNO) \cite{zhang2011parallel} generalizes GRAPPA to use a nullspace spanned by multiple annihilating kernels extracted from the ACS region.

To enable higher acceleration rates and increase flexibility in sampling pattern design, image-based  \cite{ying2007joint,trzasko2011calibrationless,holme2019enlive} and k-space-based  \cite{shin2014calibrationless,haldar2016p} calibrationless MRI (Cl-MRI) methods have been proposed and validated in research settings. Most k-space-based methods treat the recovery as a structured low-rank matrix completion problem with respect to a convolutional matrix constructed from the original k-space or weighted k-space samples. The rank deficiency results from the shift-invariant linear dependence of one k-space sample on its neighboring samples from all coils. This linear dependence originates from several sources, including multi-coil data structure \cite{griswold2002generalized,shin2014calibrationless,ying2010parallel}, limited image support \cite{shin2014calibrationless, haldar2014}, slowly varying image phase \cite{haldar2014,haldar2020linear}, multi-shot imaging \cite{mani2020improved}, piece-wise polynomial image content \cite{liang1989high,ongie2017fast}, and other transform domain sparsity \cite{jin2016general}.

Although k-space-based low-rank matrix completion methods all leverage rank deficiency, each may employ a specific optimization formulation or adopt a different optimization algorithm. For example,  Simultaneous Auto-calibrating and K-space Estimation (SAKE) \cite{shin2014calibrationless} minimizes the data mismatch subject to a hard rank constraint on the convolutional matrix using Cadzow's algorithm \cite{Cadzow1988}. Low-rank Modeling of Local K-space Neighborhoods (LORAKS) \cite{haldar2014} minimizes the sum of the squared tail singular values of the convolutional matrix with hard data consistency using the majorize-minimize (MM) algorithm \cite{haldar2014}; soft data consistency and slowly varying image phase are also optional constraints in LORAKS. For k-space based Cl-MRI methods, large memory requirement and enormous computation burden have been identified as challenges. Several algorithms aim to address these issues. For example, Generic Iterative Reweighted Annihilation Filter (GIRAF) \cite{ongie2017fast} extracts spectral information from the Gram matrix built from the convolution operator instead of directly computing a singular value decomposition of a large block-Hankel matrix.  Moreover, a ``half-circulant'' padding approximation and fast Fourier transform (FFT) increase the speed of GIRAF.  Using a synthetic ACS strategy combined with SENSE, LORAKS has been extended to 3D static imaging \cite{kim2019wave} with reduced memory and computation requirements. Likewise, through the use of slice-by-slice reconstruction in the readout direction, Annihilating Filter-Based Low Rank Hankel Matrix Approach (ALOHA) \cite{jin2016general} has been extended to 2D cine (2D+t) imaging \cite{lee2016acceleration}. Despite these computational advances, the clinical application of k-space based Cl-MRI methods is still limited by a computation complexity that is not competitive with regularized SENSE-based methods. In this work, we present a new numerical procedure, called HIgh-dimensional fast ConvolUtional framework (HICU, pronounced $`\text{h}\overline{\text{\i}} \ \text{k} \overline{\text{oo}}$) that offers these advantages: (i) small memory footprint; (ii) fast computation that is competitive with SENSE-based imaging;  (iii) seamless scalability to many MRI applications, regardless of the number of data dimensions arising from space, time, coils, or encodings; (iv) and, better reconstruction quality than SENSE-based compressed sensing (CS) reconstruction.

\section{THEORY}
A wealth of parallel imaging methods exploits the linear dependence of one k-space sample on its neighboring samples from all coils. These correlations arise both from the multi-channel interrogation of a single image using spatially smooth coil sensitivity maps \cite{griswold2002generalized,shin2014calibrationless} and from image structure itself \cite{liang1989high,haldar2014,haldar2020linear}. The dependencies yield rank deficiency in a multi-dimensional convolution matrix formed from the data, and hence the existence of annihilating kernels for the multi-dimensional k-space data. Let $\mathbb{X}$ denote the multi-dimensional array of k-space samples, and $\bm{x}$ denote the $N$-by-$1$ vectorization of the complex-valued entries in $\mathbb{X}$. For example, for a 2D+t application, the size of the four dimensional array $\mathbb{X}$ is $[ N_x,\, N_y,\, N_t,\, N_c]$, representing $N_x$ frequency encoding samples, $N_y$ phase encoding samples, $N_t$ time frames, and $N_c$ coils. Let $\mathbb{K}$ be a $\{ 0,1\}$-valued mask indicating the support of an annihilating kernel, with values on the support of $\mathbb{K}$ vectorized into the $n$-by-$1$ list $\bm{v}$. Let $\mathcal{H}_{\mathbb{K}}\{\mathbb{X}\}$ denote the $m$-by-$n$ multi-level Hankel matrix generated from the k-space array $\mathbb{X}$ such that multiplication $\mathcal{H}_{\mathbb{K}}\{\mathbb{X}\} \bm{v}$ is the vectorization of the valid convolution of $\mathbb{X}$ with multi-dimensional kernel described by $\bm{v}$. In contrast to circular and linear convolutions, valid convolution generates only those output points that do not depend on any boundary conditions. Thus, the data array $\mathbb{X}$ is lifted to a much larger, multi-level Hankel matrix, and rank of the matrix becomes a powerful implicit modeling tool \cite{haldar2020linear,JacobMagazine2020}. For accelerated imaging, only a fraction of the elements of $\mathbb{X}$ are observed; let $\mathbb{X}_0$ be the array of observed data zero-filled to match the array size of $\mathbb{X}$.  

We pose recovery of the unobserved k-space samples as a structured low-rank matrix completion problem.  
Many approaches relax the non-convex rank constraint, $\text{rank} \left\{ \mathcal{H}_{\mathbb{K}} \left(  \mathbb{X}  \right) \right\} \leq r$, to the nuclear norm minimization, which equals to minimizing the sum of all singular values 
\cite{Majumdar2012,Fazel_hankel_2013,ChenChiIT,otazo2015low,mani2017}. 
Instead, we choose as a cost function the squared distance
of the completed multi-level Hankel matrix to the set of matrices with rank not exceeding $r$:
\begin{equation}
    \label{eq:l2tail}
    J( {\bm{x}} )  = \sum_{k > r}  \sigma^2_k \left\{ \mathcal{H}_{\mathbb{K}} \left(  \mathbb{X}  \right) \right\} ,
\end{equation}
where $\sigma_k \{ {\bm{A}} \}$ denotes the $k^{\text{th}}$ largest singular value of a matrix $\bm{A}$.
This cost function was introduced in the LORAKS framework \cite{haldar2014} and is the sum of squares of the tail singular values, $\sigma_k$ for $k>r$,
of the multi-level Hankel matrix formed from the k-space array,  $\mathbb{X}$. 
This cost may be minimized subject to constraints, $f_i ({\bm{x}}) = 0$, $i=1, ... , c$. An example constraint is data fidelity, $f_1 ( {\bm{x}})  = \| \mathbb{M} \circ \mathbb{X} - \mathbb{X}_0\|\frob^2$, where $\mathbb{M}$ is the $\{ 0,1\}$-valued mask indicating observed k-space locations, $\circ$ denotes element-wise multiplication, and $\| \cdot \|\frob$ denotes the Frobenius norm. 
In general, the constrained minimizer, ${\bm{x}}_{\star}$, yields $J({\bm{x}}_{\star})>0$, and the matrix completion is only approximately rank $r$.

Given annihilating kernel mask $\mathbb{K}$, we adopt for the $m$-by-$n$ multi-level Hankel matrix the shorthand notation ${\bm{H}}({\bm{x}}) = \mathcal{H}_{\mathbb{K}}\left( \mathbb{X} \right) $, and use ${\bm{H}}\herm({\bm{x}})$ to denote the conjugate transpose.   The cost, $J( {\bm{x}} )$, in \eqref{eq:l2tail} is readily seen to be equivalent to
$ \min_{ {\bm{Q}}} \sum_{j=r+1}^{n} \|  {\bm{H}}({\bm{x}}) {\bm{Q}} \|\frob^2$, with ${\bm{Q}}\herm {\bm{Q}} = {\bm{I}}$.
Each of the $n-r$ orthonormal columns of $\bm{Q}$ specifies an approximate
annihilating filter, and $J(\bm{x})$ is the residual energy, after valid convolution, summed across all $n-r$ approximate annihilating kernels. Thus, in HICU the matrix completion task is formulated as the optimization
\begin{equation}
    \label{eq:hicu_cost2}
    \min_{{\bm{x}}, {\bm{Q}}} \ \|{\bm{H}}({\bm{x}}) {\bm{Q}}  \|\frob^2 + \sum_{i=1}^{c} \lambda_i f_i({\bm{x}}) , \quad {\bm{Q}}\herm {\bm{Q}} = {\bm{I}},
\end{equation}
with Lagrange multipliers, $\lambda_i$ \cite{haldartechreport}. We adopt the familiar alternating minimization strategy for optimizing
over $\bm{x}$ and $\bm{Q}$: first update the nullspace basis, $\bm{Q}$, then update the unobserved k-space samples, $\bm{x}$. We demonstrate that this cost function pairs quite amicably with four numerical techniques that, taken together, yield both fast computation and low memory requirement.  The proposed iteration, dubbed HICU, is given in Algorithm~\ref{alg:h2_new}. 
We note that by 
choosing an annihilating kernel of size $N_c$ in the coil dimension, valid convolution affects a sum across all coils, while maintaining the conceptual simplicity of a single multi-dimensional convolution.

Consider now the four numerical techniques we adopt in Algorithm~\ref{alg:h2_new}. First, we use randomization
to efficiently compute a nullspace basis.
To this end, we use the randomized singular value decomposition (rSVD) \cite{Tropp2011}
to compute the $n$-by-$r$ matrix, ${\bm{V}}^{(i)}$, corresponding to the principal $r$ right singular values 
of ${\bm{H}} ( {\bm{x}}^{(i-1)} )$; the update to ${\bm{Q}}^{(i)}$ is then completed from ${\bm{V}}^{(i)}$ using $r$ Householder reflections. So, ${\bm{Q}}^{(i)}$ minimizes [\ref{eq:hicu_cost2}] subject to ${\bm{Q}}\herm  {\bm{Q}} = {\bm{I}}$ for fixed ${\bm{x}}^{(i-1)}$.  
The rSVD can compute the $r$ principal right singular vectors, ${\bm{V}}^{(i)}$, to high precision using only $\frac{3}{2}r$ applications of the multi-dimensional valid convolution, ${\bm{H}} ( {\bm{x}}^{(i-1)} )$, and its conjugate transpose, 
${\bm{H}}\herm ( {\bm{x}}^{(i-1)} )$, which is itself a valid convolution operation. Thus, rSVD avoids explicit construction of both the large multi-level Hankel matrix and its Gram matrix, sidestepping the associated memory requirement and computation time. Compared to other truncated SVD techniques, such as Lanczos bidiagonalization algorithm \cite{larsen1998lanczos} found in PROPACK \cite{larsen2004propack}, rSVD is more stable and inherently amenable to parallel computing \cite{Tropp2011}. Recent studies have shown that rSVD methods can be significantly faster compared to the truncated SVD implementation in PROPACK \cite{feng2018faster}.

Next, we make a number, $G^{(i)}$, of gradient descent steps on $\bm{x}$ for fixed ${\bm{Q}}^{(i)}$. A gradient descent step updates the unobserved k-space points to reduce the energy in the set of annihilator outputs, as constrained by the costs $\lambda_i f_i({\bm{x}})$.  The number of steps can be chosen to balance computational cost between the rSVD to update $\bm{Q}^{(i)}$ and the total collection of $G^{(i)}$ gradient descent steps to update ${\bm{x}}^{(i)}$. Only a few gradient steps are required, and computation does not benefit from full convergence for a fixed nullspace estimate, ${\bm{Q}}^{(i)}$.

Exact computation of the gradient of $\| {\bm{H}} ({\bm{x}}^{(i-1)}) {\bm{Q}}^{(i)} \|\frob^2$ for fixed ${\bm{Q}}^{(i)}$ requires a sum over all annihilating kernels of the composition of two convolutions. So, at each gradient step we adopt a second numerical technique for speed while preserving the correct mean gradient direction: motivated by the random mixing of dimension reduction in the Johnson-Lindenstrauss (JL) Lemma \cite{JL1984,BaraniukJL}, we reduce ${\bm{Q}}^{(i)}$ to $p$ columns. Specifically, before computing a gradient direction, we use an $n$-by-$p$ matrix, ${\bm{P}}^{(i,j)}$, of i.i.d.\ zero-mean Gaussian entries with variance $1/p$ to project ${\bm{Q}}^{(i)}$ to a smaller $p$-dimensional nullspace, $\widetilde{\bm{Q}}^{(i,j)}={\bm{Q}}^{(i)} {\bm{P}}^{(i,j)}$. 
Compared to randomly sampling $p$ columns from ${\bm{Q}}^{(i)}$,
the JL compression 
provides implicit preconditioning. 

Third, exact line search (ELS) can be efficiently implemented for $J({\bm{x}})$ paired with many choices of $f_i$, providing the minimum of the cost function in [\ref{eq:hicu_cost2}] in the direction of the negative of the gradient and obviating any need for a step-size heuristic or iterative line search.  The step size at Step 9 of Algorithm~\ref{alg:h2_new} is found via one convolution with each of the $p$ small annihilating kernels in $\widetilde{{\bm{Q}}}^{(i,j)}$. 

Fourth, the region of k-space, $\mathbb{S}^{(i)}$, on which the valid convolution
is computed in Steps 2 and 8 does not affect the $n$-by-$r$ matrix of principal right singular vectors in the idealized noiseless case. Therefore, invoking shift invariance across the spatial frequency dimensions in k-space, the region may be restricted to small subsets of the full k-space.  In this manner, a high signal-to-noise ratio (SNR), under-sampled, small region at the center of k-space may be used to rapidly determine an approximate solution
for ${\bm{Q}}^{(i)}$.  Subsequent iterations can use more or all of the k-space to both refine ${\bm{Q}}^{(i)}$ and estimate the full k-space data array, $\mathbb{X}$. 
We refer to this strategy as center-out (CO). A similar use of the center of k-space is found,
for example, in SAKE \cite[p.~968]{shin2014calibrationless} and Wave-LORAKS \cite{kim2019wave} where an auto-calibration region is synthesized as a pre-processing step to a SENSE reconstruction. In Algorithm~\ref{alg:h2_new}, the abbreviated notation ${\bm{H}}^{(i)}\left( {\bm{x}}^{(i-1)} \right)$ is used to denote this potentially much smaller multi-level Hankel matrix constructed using a portion, $\mathbb{S}^{(i)}$, of the full k-space at iteration $i$. 

Let $s=|\mathbb{S}^{(i)}|$ be the number of output points from the valid convolution region at iteration $i$. Recall $n=| \mathbb{K}|$ is the number of convolution kernel samples, and $r$ is the rank;  note $r < n < s$. The complexity of gradient computation with ELS is roughly $3nps$, and the rSVD complexity is roughly $3nrs$. Note that $r$ generally grows with number of dimensions in the data array, $\mathbb{X}$. Thus, for high dimensional problems, the ratio $r/p$ grows large, leaving the rSVD much more costly than a single gradient descent step. The memory requirement is approximately $N$, the size of k-space array $\mathbb{X}$, plus $1.5rs$.

The combination of the optimization objective \eqref{eq:hicu_cost2} and four numerical techniques adopted here---CO, rSVD, JL, ELS---
provides significant savings in computation time for memory-efficient completion of a multi-level Hankel matrix with approximate rank deficiency.
The moniker ``HIgh-dimensional fast ConvolUtional framework'' refers to the simple and exclusive reliance in Algorithm~\ref{alg:h2_new} on multi-dimensional convolutions, ${\bm{H}}^{(i)} \left( {\bm{x}}^{(i-1)} \right)$ 
and ${\bm{H}}^{(i) \prime} \left( {\bm{x}}^{(i-1)} \right)$, at each step of the alternating minimization, and the name points to the attendant scalability of memory and computation to large k-space arrays, $\mathbb{X}$.

Below, we present numerical results for HICU computation for the case of exact data matching; this allows direct comparison with publicly available code for SAKE \cite{shin2014calibrationless,bart-toolbox} and coincides with one specific cost function found in the LORAKS 2.1 suite \cite{haldartechreport}, namely C-based LORAKS with hard data constraint. To achieve data matching in HICU, we simply set, in Step 8 of Algorithm~\ref{alg:h2_new}, the gradient of $J(\bm{x})$ equal to zero at observed k-space locations.

\section{METHODS}
We evaluated HICU using four in vivo studies: 2D static imaging of the brain (Study I), 2D+t imaging of the heart (Study II), 3D imaging of the knee (Study III), and MSDWI of the brain (Study IV). 

\subsection{MR acquisition}    
\begin{figure}[!ht]
    \centering
    \includegraphics[width = 16.5cm]{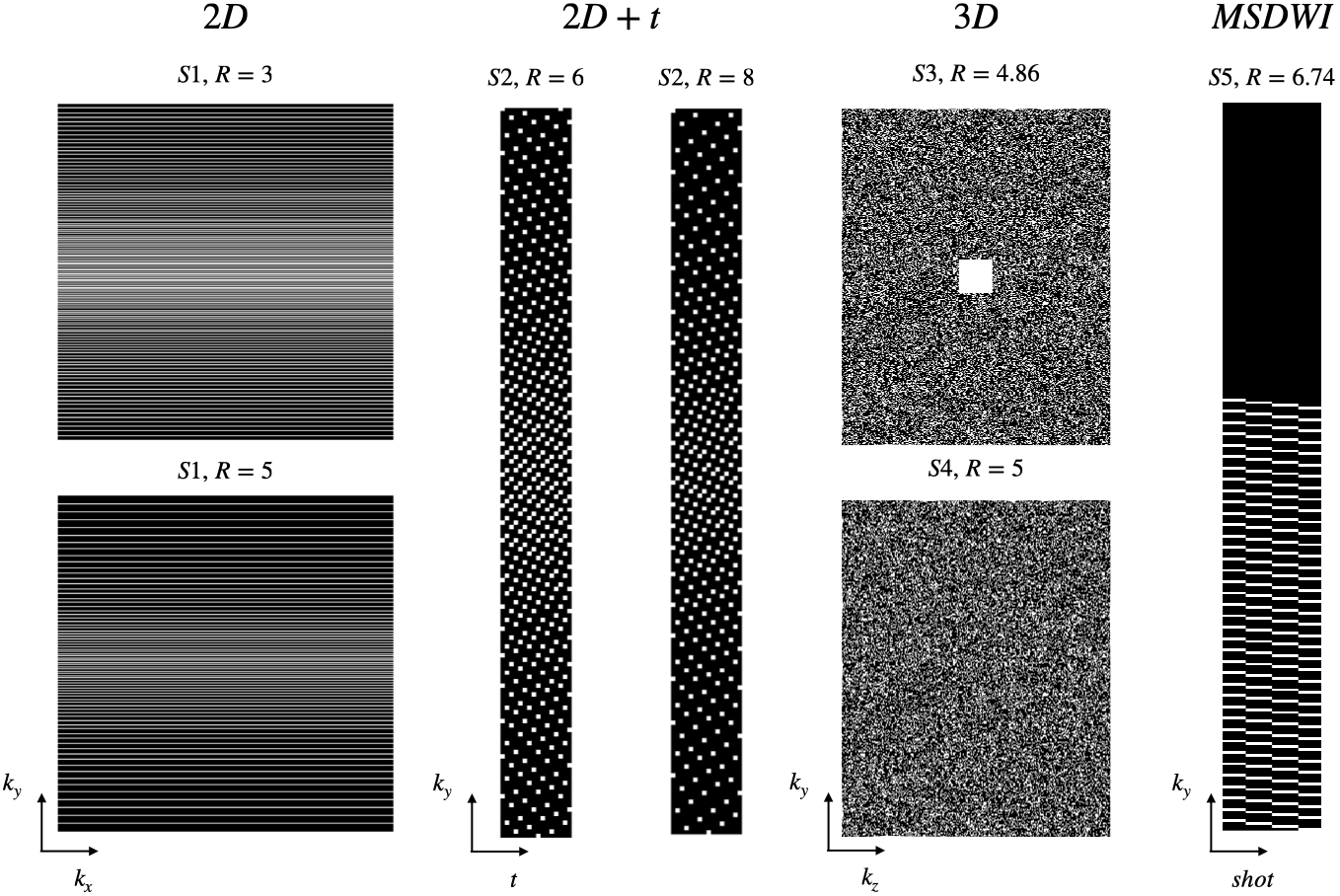}
    \caption{Sampling patterns ($S1$ to $S5$) and acceleration rates used in the 2D, 2D+t, 3D, MSDWI studies.}
    \label{Figure_nD_Sampling_Pattern}
\end{figure}

In Study I, six T2-weighted pMRI brain datasets, $B1, B2, \cdots, B6$, were taken from the New York University fastMRI database \cite{zbontar2018fastmri}. All datasets were fully sampled and collected in an axial orientation on $3$\,T scanners using a T2-weighted turbo spin-echo sequence. Other imaging parameters included: TE 113~ms, TR 6,000---6,570~ms, TI 100~ms, field-of-view (FOV) 220~mm$ \times$ 227~mm, slice thickness 7.5~mm,  matrix size 384$\times$384, number of receive coils 16---20, and flip angle 149---180 degrees. The center slice from each dataset was used. The 2D datasets were retrospectively undersampled using a variable density sampling pattern, $S1$, at rates $R=3$ and $R=5$, as shown in Figure~\ref{Figure_nD_Sampling_Pattern}. The pattern $S1$ had at most two contiguous readouts at $R=3$ and no contiguous readouts at $R=5$.

In Study II, seven fully sampled 2D+t cardiac datasets $C1, C2, \cdots, C7$ were collected from seven healthy volunteers using a balanced steady-state free precession sequence under breath-hold with prospective triggering.
The data were collected at The Ohio State University and Nationwide Children's Hospital, with the ethical approval for recruitment and consent given by an Internal Review Board (2005H0124). Three short-axis oriented and four long-axis oriented fully sampled segmented datasets were collected on $3$\,T scanners (MAGNETOM Prisma, Siemens Healthineers, Erlangen, Germany). Other imaging parameters included: TE 1.48---1.53~ms, TR 2.95---3.05~ms, FOV 380---400~mm $\times$ 285---325~mm, slice thickness 8~mm, matrix size 176---384 $\times$ 132---156, number of frames 16---25, temporal resolution 36.6---38.35~ms, number of receive coils 20---34, and flip angle 33---42 degrees. The 2D+t datasets were retrospectively undersampled at $R=6$ and $R=8$ using a variable density pseudo-random sampling pattern $S2$ shown in Figure~\ref{Figure_nD_Sampling_Pattern}. ALOHA reconstruction requires sampling of the center phase encoding line; therefore, for the ALOHA method only we added an additional line to the sampling pattern for each frame, resulting in slightly lower acceleration rates of $R= 5.77$ and $R=7.61$. 

In Study III, five 3\,T 3D knee datasets, $D1, D2, \cdots, D5$ from www.mridata.org were used. The spin-echo imaging parameters included: receiver bandwidth 50~kHz, number of coils $N_c=8$, FOV 160 mm $\times$ 160 mm $\times$ 153.6 mm, matrix size $320 \times 320 \times 256$, repetition time 1,550~ms, echo time $25$~ms, and flip angle~$90$\,degrees. The datasets were retrospectively undersampled along the phase encoding dimensions using 2D random sampling patterns $S3$ and $S4$ shown in Figure~\ref{Figure_nD_Sampling_Pattern}. Pattern $S4$ is 2D random sampling with rate $R=5$, while $S3$ augments $S4$ with a $32 \times 32$ fully sampled center region to yield $R=4.86$.

In Study IV, one ten-slice dataset was provided courtesy of the University of Iowa. The data were acquired from a healthy volunteer on a 3~T GE Discovery MR750W (GE Healthcare, Waukesha) using a four-shot ($N_s=4$) dual spin echo diffusion sequence. Parameters included: partial Fourier $59\%$,  TE $84$ ms for $b$-value of $700~\text{s/mm}^2$, FOV $210 \times 210$\,mm, sampling matrix $256 \times 152$, slice thickness $4$\,mm, slice number $10$, $\text{NEX}=2$, $32$ coils, one non-diffusion image, and $60$ diffusion directions. The prospective sampling pattern for four shots is $S5$ with acceleration rate $R=6.74$, as shown in Figure~\ref{Figure_nD_Sampling_Pattern}. Detailed description of the dataset is provided in~\cite{mani2020improved}

\subsection{MR reconstruction and analysis}
In Study I, the datasets were compressed to $N_c=8$ virtual coils before reconstruction \cite{zhang2013coil}. For comparison, we include three reconstruction methods: SAKE using the authors' publicly available Matlab code \cite{SPIRiTV(0.3New):2020},
LORAKS 2.1 using the authors' publicly available Matlab code \cite{haldartechreport};
and, Image Reconstruction by Regularized Nonlinear Inversion (NLINV)  \cite{uecker2008image} using compiled C code from the Berkeley Advanced Reconstruction Toolbox (BART) \cite{bart-toolbox}. To allow uniform comparison to SAKE, the C-based version of LORAKS was used, and kernels for HICU and LORAKS were restricted to rectangular support of size $[5,5, N_c]$. The sixth dataset, $B6$ at ($S1, R=5$), was used to tune the rank for SAKE, LORAKS, and HICU manually to maximize reconstruction signal-to-error ratio (SER). SER is defined as $\text{SER} = 20\log ( \|\mathbb{X}\|\frob / \| \hat{\mathbb{X}}-\mathbb{X}\|\frob)$ \cite{erdogmus2004image}. The remaining five datasets were then used for performance evaluation at one hour execution time.  Coincidentally, SAKE, LORAKS, and HICU all shared the same best rank, $r=30$. For LORAKS, algorithm choices were set for consistent comparison to SAKE, i.e., hard data consistency and no virtual conjugate coils. Additionally, the multiplicative half-quadratic algorithm using FFT approximation was chosen with LORAKS for execution speed. From computed results, the SER versus time for SAKE, LORAKS, and HICU was averaged over $B1, B2, \cdots, B5$. For NLINV, the input zero-filled k-space was scaled to have the Frobenius norm equal to $100$. 
The number of iterations for NLINV was tuned to maximize the SER for ($B6,S1,R=3$) and ($B6,S1,R=5$). 
SER for NLINV at each iteration is not available from the compiled code; thus, only the final SER is reported for NLINV. For the first stage of HICU, 
the size of the CO region was set at $\frac{1}{4}N_x \times \frac{1}{4}N_y \times N_c$, with $p=N_c=8$ and $G^{(i)} = 5$ gradient steps per update of the nullspace.  For the second stage, the full k-space was used with $p = 4N_c=32$ and $G^{(i)} = 10$. From the tuning results, the number of iterations for the first stage was set at $50$ for $R=3$ and $200$ for $R=5$. Coil images were combined via square-root sum of squares (SSoS) \cite{larsson2003snr}. 

For all four methods, the mean and standard deviation were computed for four metrics: k-space SER in dB, high frequency error norm (HFEN) \cite{ravishankar2010mr} in dB, structural similarity index (SSIM), which was averaged across all coil magnitude images, and the time, T$_c$, to reach within $0.1$\,dB of the SER achieved at one hour (except for NLINV). For NLINV, T$_c$ corresponds to the computation time to run $14,\,15$ iterations for $R=3,\,5$. To illustrate the effect of $p$ in HICU, the SER curves for ($B1, S1, R=3$) were computed for six values of $p$ ranging from $1$ to $n-r=170$. To illuminate the separate and joint effects of CO and JL strategies in HICU, the SER curves for ($B1, S1, R=5$) were computed for all four combinations of using or omitting the strategies.

In Study II, the data were compressed to $N_c=12$ virtual coils before reconstruction. For comparison, we included four reconstruction methods: ALOHA \cite{lee2016acceleration, ALOHA}; total-variation (TV) penalty using MFISTA \cite{tan2014smoothing}; soft-thresholding of non-decimated wavelet transform (NWT) using balanced FISTA \cite{ting2017fast}; and low-rank plus sparse (L+S) \cite{otazo2015low, L+S2018}. The seventh dataset $C7$ at ($S3, R=8$) was withheld to determine the tuning parameters. For all methods, the initialization was the time-averaged k-space; if a phase encoding point was not sampled across all frames, then zero filling was used. For TV, NWT, and L+S, the number of iterations was $150$, and
the sensitivity maps were estimated from time-averaged k-space using the method by Walsh et al.\ \cite{walsh2000adaptive}. The number of iterations and tolerance for two stages of ALOHA were set to $[500,\, 500]$ and $[10^{-2},\, 10^{-3}]$; the ALOHA kernel size was $[5,\, 5]$. For HICU, the kernel size was $\left[ 5,\,5,\,5,\,N_c\right]$, rank $r=130$, $p = N_c = 12$, and the total number of iterations $I=101$. For the first $100$ HICU iterations, $\mathbb{S}^{(i)}$ was the center $\frac{1}{4}N_x \times \frac{1}{4}N_y \times N_t \times N_c$ and $G^{(i)}=5$;  $\mathbb{S}^{(101)}$ was the full k-space and $G^{(101)} = 100$. In HICU computation, the convolution in the time dimension was circular, rather than valid. We consistently observed $0.2$ to $0.3$\,dB SER gain with circular convolution along time, compared to valid convolution. For the three SENSE-based methods, the reconstruction results were converted into coil images via pixel-wise multiplication of the reconstructed image with the sensitivity maps. For all five methods, the mean and standard deviation were computed for four metrics: k-space SER, HFEN, SSIM, and compute time.

In Study III, we compared HICU to two reconstruction methods: SENSE-based reconstruction with regularization of wavelet transform (WT) sparsity using BART \cite{bart-toolbox}; and, NLINV using BART \cite{bart-toolbox}. The fifth dataset $D5$ was withheld to determine the tuning parameters. For HICU and NLINV, the parameters were based on sampling pattern $S5$ and used for both $S4$ and $S5$. For WT, one set of coil sensitivity maps was extracted using ESPIRiT \cite{uecker2014espirit}; parameters were $\lambda=0.01$ and $150$ FISTA iterations. For NLINV, the observed k-space was scaled to have Frobenius norm $25,600$, and the number of iterations was set at $18$. For HICU, the kernel size was $\left[ 5,\,5,\,5,\,N_c\right]$, rank $r=150$, $p = N_c = 8$, and $I=11$. For the first ten iterations, $\mathbb{S}^{(i)}$ was $\frac{1}{4}N_x \times \frac{1}{4}N_y \times \frac{1}{4}N_z \times N_c$ and $G^{(i)}=5$. $\mathbb{S}^{(11)}$ was the full k-space, and $G^{(11)}=20$. For all five methods, the mean and standard deviation were computed for four metrics: k-space SER, HFEN, SSIM, and computing time.

In Study IV, reconstruction for each slice was performed separately after compressing to $N_c = 8$ virtual coils. IRLS MUSSELS with conjugate symmetry  \cite{mani2020improved} was used for comparison. For IRLS MUSSELS, the coil sensitivity map was extracted using ESPIRiT  \cite{uecker2014espirit} based on the sum of the $4$ shots for the non-diffusion image. The kernel size was $[5,\,5]$; execution employed $10$ outer iterations and $8$ inner iterations, with regularization parameter $\lambda = 0.005$. HICU was applied separately for each $b$ value because the phase modulation due to multi-shot varies with diffusion weighting. To better accommodate the uniform downsampling, the nullspace was augmented $[{\bf Q}\,|\,{\bf Q}_+]$ in Step 3 of Algorithm \ref{alg:h2_new} using information from the non-diffusion ($b0$) measurement. Forty null vectors, ${\bf Q}_0$, were computed using the sum of the four shots at $b0$.  Then, ${\bf Q}_+$ is a block diagonal matrix, with four repetitions of ${\bf Q}_0$  along the diagonal.
For HICU, the kernel size was $\left[ 5,\,5,\,N_c,\,N_s\right]$, rank $r=45$, $p = N_c = 8$, and $I=51$. For the first 50 iterations, $\mathbb{S}^{(i)}$ was center $\frac{1}{4}N_x \times \frac{1}{4}N_y \times N_c \times N_s$ and  $G^{(i)}=5$;
$\mathbb{S}^{(51)}$ was the partial Fourier sampled region of k-space, with $G^{(51)} = 100$. HICU matrix completion was followed by homodyne partial Fourier reconstruction \cite{noll1991homodyne}. 
In absence of fully sampled data, no quantitative metrics were computed.

For all studies, processing was conducted in Matlab 2020a (Mathworks, Natick, MA, US) on an Alienware Aurora Ryzen\texttrademark~desktop, equipped with an AMD Ryzen 3950X CPU and 64\,GB memory. The Matlab code for HICU is provided at \url{https://github.com/OSU-CMR/HICU}.
    
\section{RESULTS}
\begin{figure}[!ht]
    \centering
    \includegraphics[width = 16.5cm]{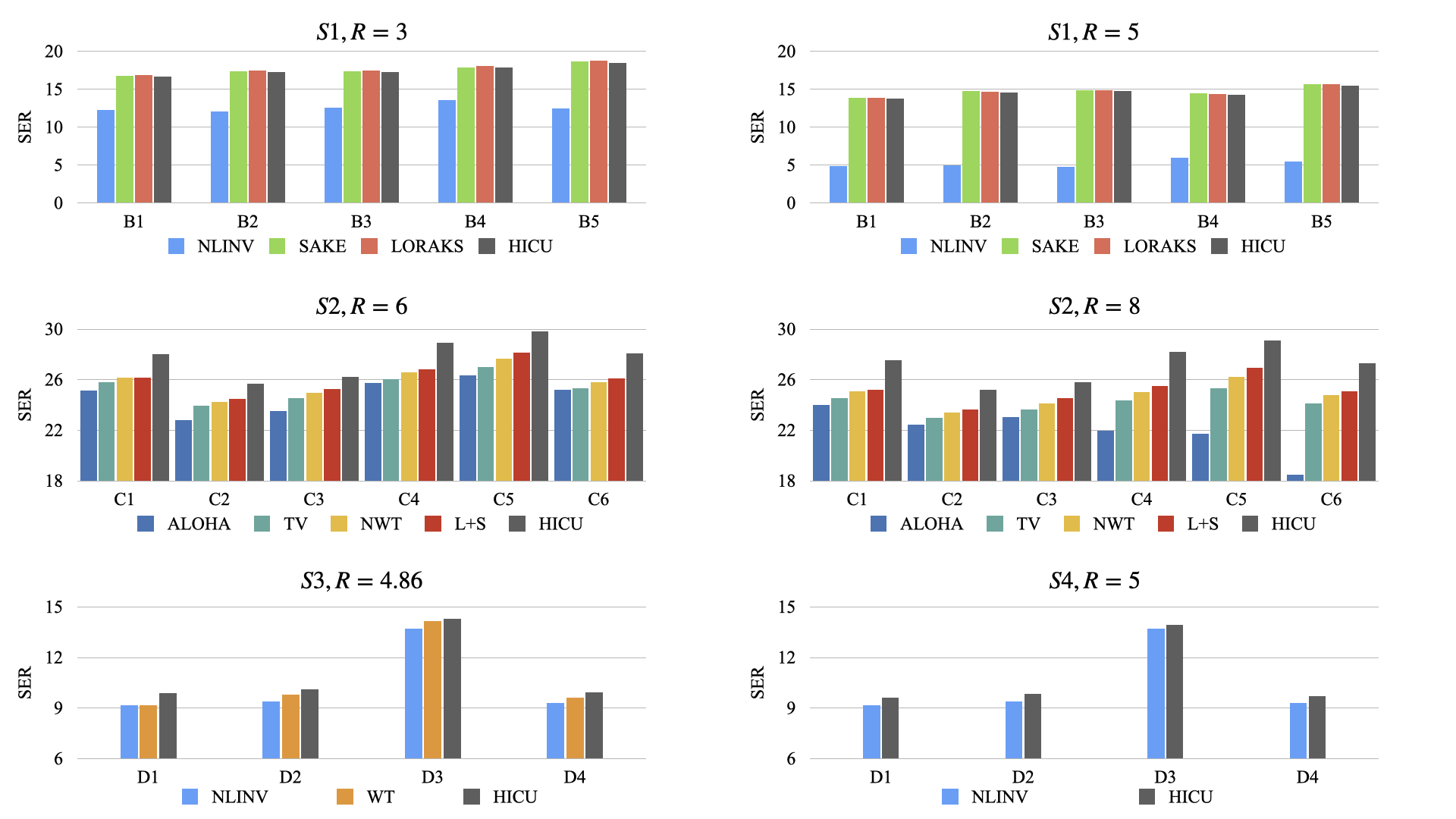}
    \caption{SER of reconstruction results from all methods and for all datasets and sampling patterns.}
    \label{Figure_nD_SER}
\end{figure}

For the first three studies, Table \ref{Tab:nD SER SSIM HFEN} reports the mean and standard deviation of quantitative metrics for all datasets, sampling patterns, and accelerations. Figure \ref{Figure_nD_SER} shows the SER for each dataset.

\begin{figure}[!ht]
    \centering
    \includegraphics[width = 16.5cm]{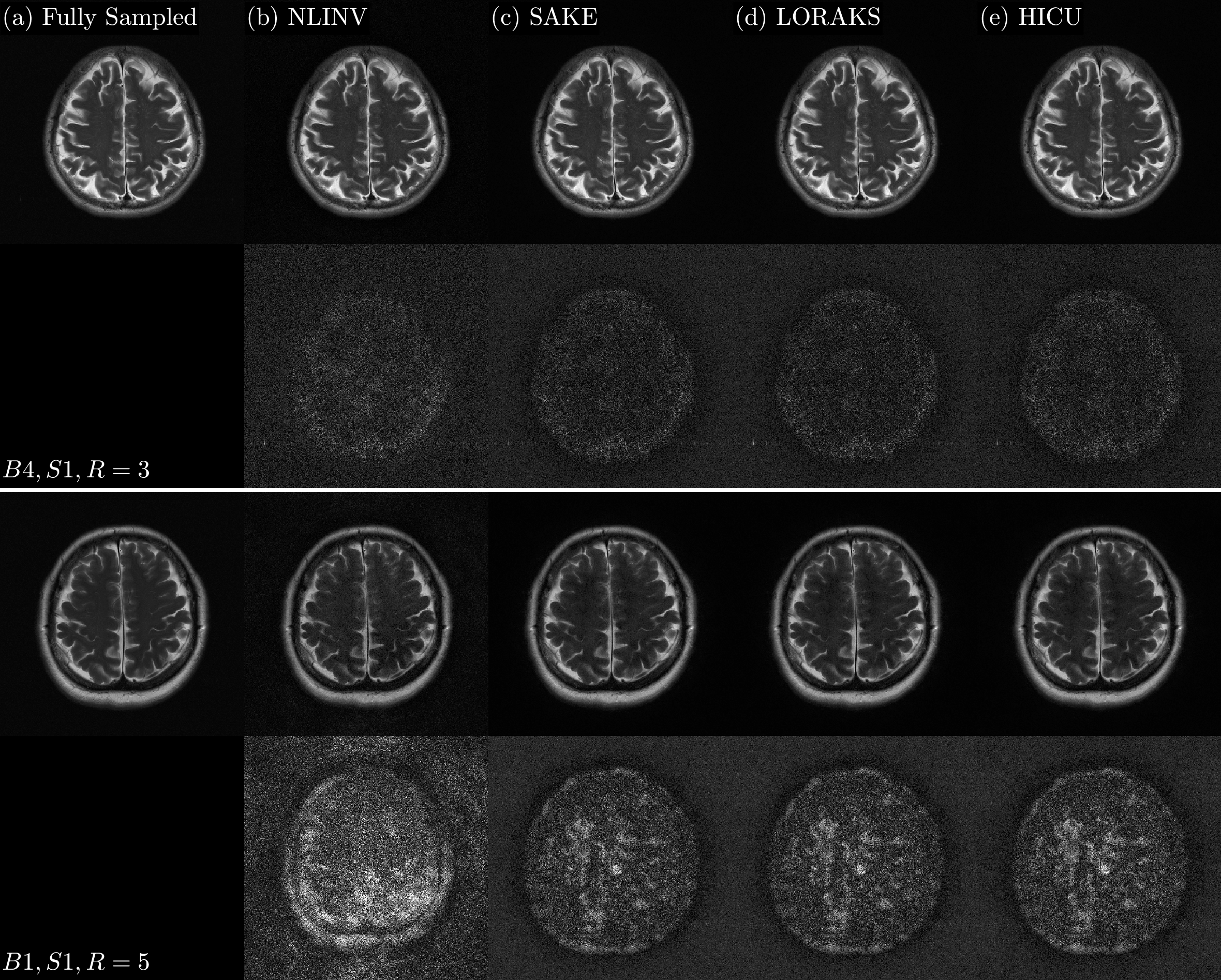}
    \caption{Top row: SSoS images for SAKE, LORAKS, HICU at one hour, and NLINV at the final iteration. Bottom row: $10\,\times$\,absolute error relative to fully sampled k-space.}
    \label{Figure_2D_Images}
\end{figure}

For Study I, the memory requirements for SAKE, LORAKS, NLINV, and HICU were approximately $450$\,MB, $450$\,MB, $130$\,MB,  and $18$\,MB, respectively. Figure \ref{Figure_2D_SER_vs_Time} shows the average SER versus runtime  for SAKE, LORAKS, and HICU for ($S1,R=3$) and ($S1,R=5$).  Time in seconds is plotted logarithmically to illuminate computing time ratios and simultaneously view time on the scale of hours, minutes, and seconds. Figure \ref{Figure_2D_All_Module_Effect} shows SER versus runtime for six choices of $p$ with CO disabled to explore the effect of the JL projection dimension, $p$, for ($B1,S1,R=3$); this figure also shows SER versus runtime for four HICU variants with and without the CO and JL numerical strategies for ($B1,S1,R=5$). Figure \ref{Figure_2D_Images} shows the SSoS reconstruction images for ($B4,S1,R=3$) and ($B1,S1,R=5$), the two datasets that yielded the highest and lowest SER from Figure~\ref{Figure_nD_SER}. 

For Study II reported in Table~\ref{Tab:nD SER SSIM HFEN}, HICU yields the highest average SER, SSIM, and HFEN. For display of representative images, we selected the two combinations of dataset, sampling pattern, and acceleration rate that yielded the highest and lowest SER values in Figure~\ref{Figure_nD_SER}. 
Videos of all frames and error maps for ($C5, S3, R=6$, $C2, S3, R=8$) are found in the Supporting Information Videos S\ref{video 1} and S\ref{video 2}.

\begin{figure}[!ht]
    \centering
    \includegraphics[width = 16.5cm]{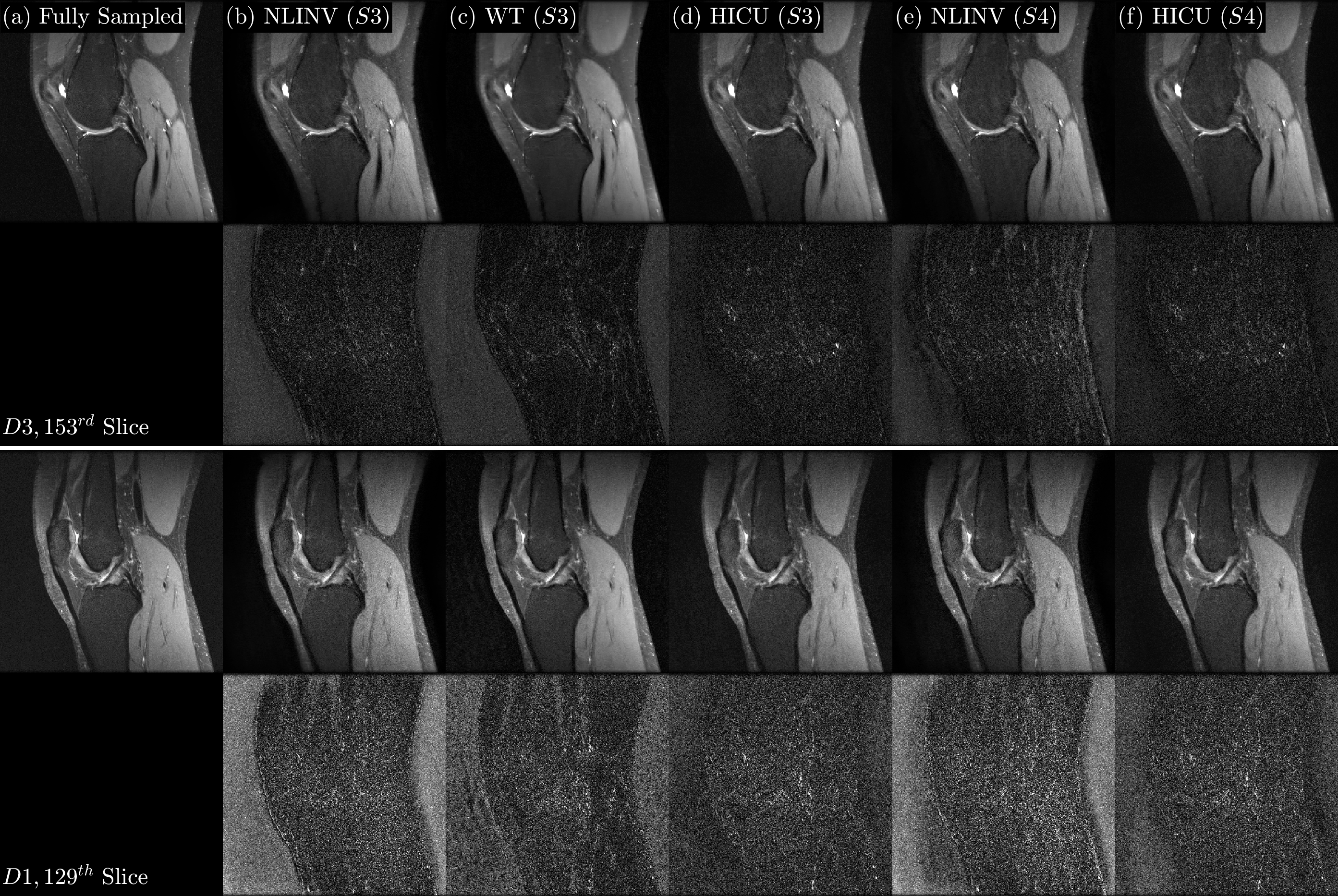}
    \caption{Top: SSoS image for all methods of $153^{\text{rd}}$ slice of $D3$ and $129^{\text{th}}$ slice of $D1$. Bottom: $7\,\times$\,absolute error.}
    \label{Figure_3D_Images}
\end{figure}

\begin{figure}[!ht]
    \centering
    \includegraphics[width = 16.5cm]{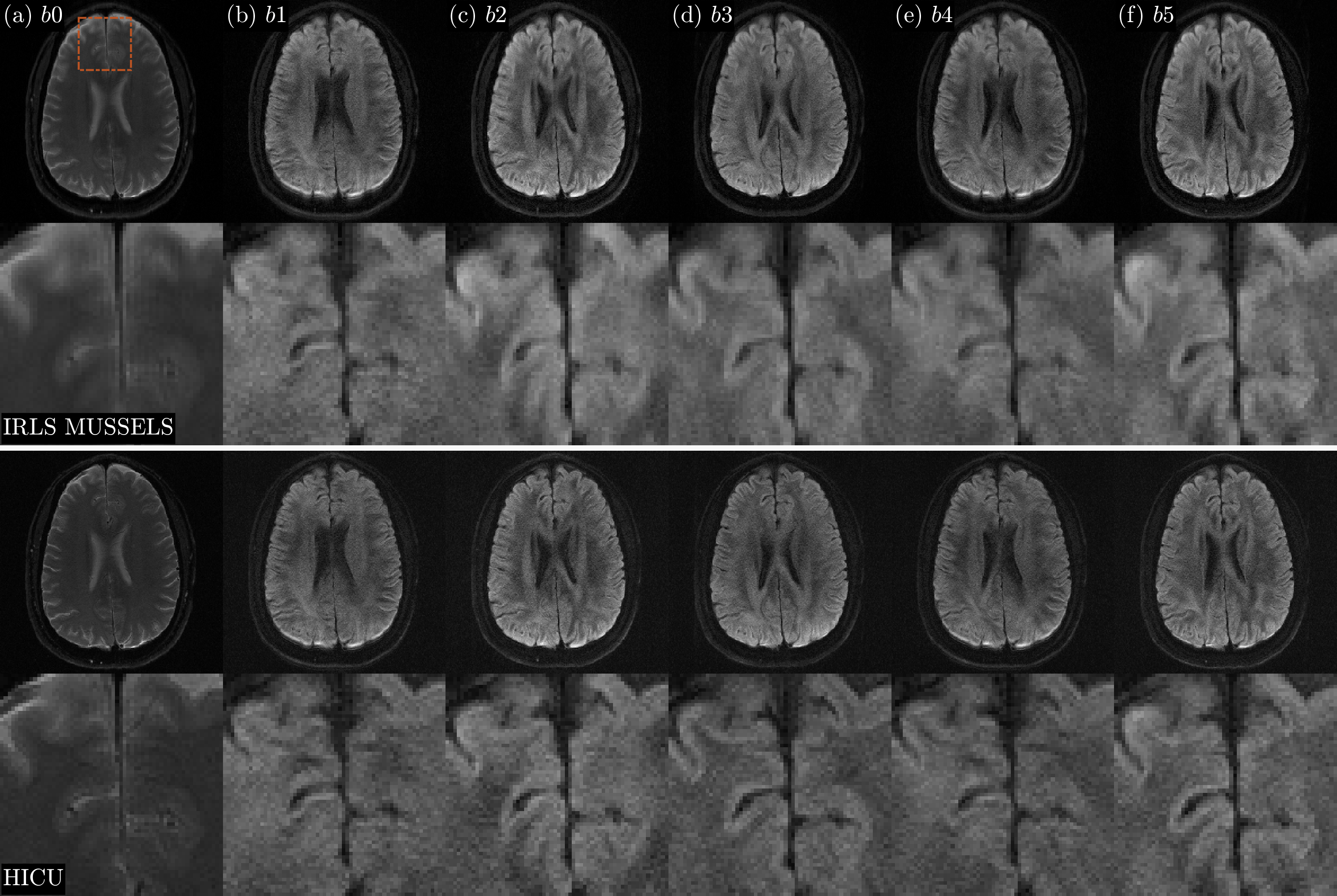}
    \caption{SSoS image for HICU and IRLS MUSSELS with conjugate symmetry for $b0,\cdots, b5$. Results from the 6$^\text{th}$ slice are shown.}
    \label{Figure_MSDWI_Image}
\end{figure}

For Study III reported in Table~\ref{Tab:nD SER SSIM HFEN}, HICU yields the highest average SER, SSIM, and HFEN. Figure \ref{Figure_3D_Images} shows the reconstruction results from $D3$ and $D1$, which had the
highest and lowest average k-space SER in Figure~\ref{Figure_nD_SER}. For Study IV, Figure~\ref{Figure_MSDWI_Image} shows the SSoS of all coils and shots for the first six $b$ values. The results from the 6$^\text{th}$ slice are shown. From Table \ref{Tab:nD SER SSIM HFEN}, HICU is over $5$ times faster than SENSE-based IRLS MUSSELS with conjugate symmetry. 

\section*{DISCUSSION}
\begin{figure}[!ht]
    \centering
    \includegraphics[width = 8.5cm]{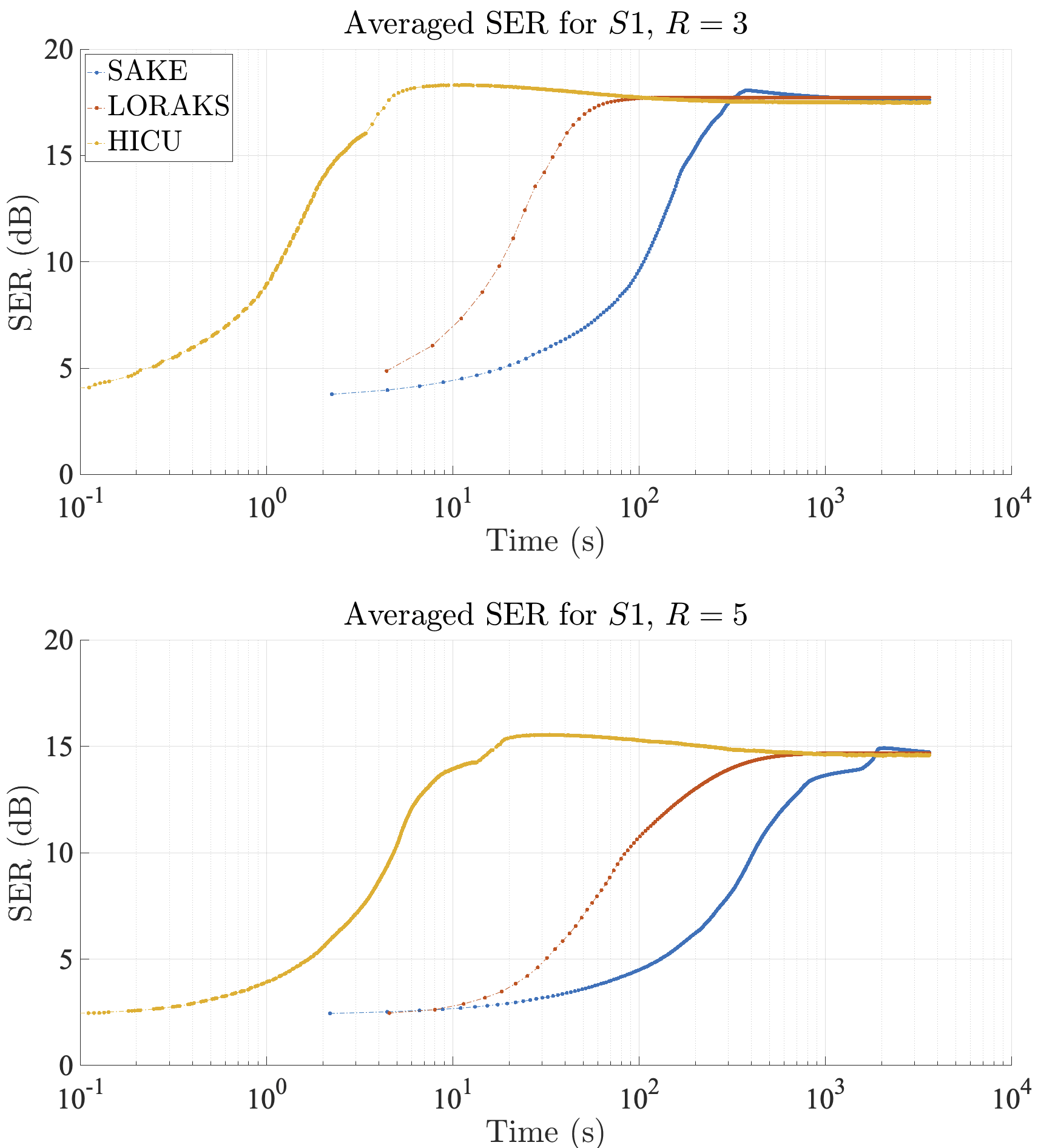}
    \caption{SER versus runtime for $S1,R=3$ and $S1,R=5$. Each curve represents SER values that are averaged over five datasets.}
    \label{Figure_2D_SER_vs_Time}
\end{figure}

HICU provides a fast and memory-efficient computational procedure for structured low-rank matrix completion. Using 2D brain images, Study I suggests, from Table \ref{Tab:nD SER SSIM HFEN} and Figure~\ref{Figure_2D_SER_vs_Time}, that HICU can provide over an order of magnitude speedup compared to SAKE and LORAKS while providing comparable image quality across all metrics and significantly reduced memory requirement. Likewise, Figure~\ref{Figure_2D_Images} shows qualitatively very similar converged reconstruction results for SAKE, LORAKS, and HICU; the similarity is expected in that all three methods exploit the rank deficiency of the Hankel data matrix. The annihilating kernel support used in Study I was rectangular to conform with SAKE implementation \cite{bart-toolbox}; however, LORAKS and HICU can support circumscribing circular or spherical kernel masks that have been observed to yield SER gains \cite{haldar2014}. NLINV produced much lower image quality in this application than the other methods and required $1.5$ to $4$ times more computation time than HICU.

From Figure~\ref{Figure_2D_SER_vs_Time}, SAKE and HICU exhibit a decline in SER as iterations proceed past the point of maximum SER.  This has been previously observed \cite{lustig2010spirit} and may be attributed to the high SER central k-space region used in the CO strategy.  High SER of the nullspace can lead to noise amplification in the recovered k-space; this seemingly paradoxical effect has been studied for GRAPPA \cite{DingParadox2015}. The presence of a time dimension in the 2D+t images of Study II ameliorates this behavior.

\begin{figure}[!ht]
    \centering
    \includegraphics[width = 8.5cm]{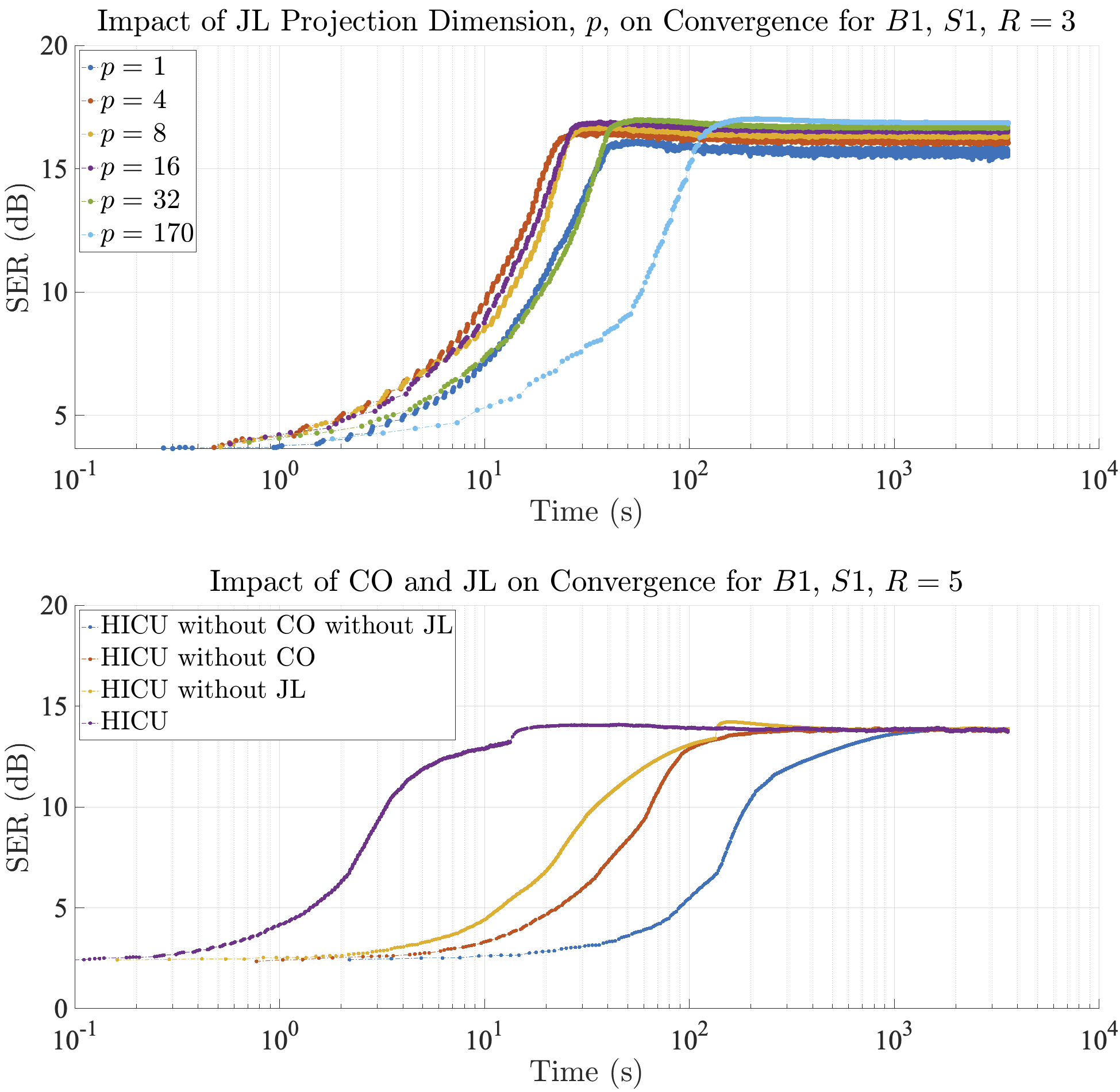}
    \caption{(top) The effect of JL on convergence speed for $p = 1,4,8,16,32,170$ for ($B1,S1,R=3$). (bottom) The impact of CO and JL on convergence speed for ($B1,S1,R=5$).}
    \label{Figure_2D_All_Module_Effect}
\end{figure}

Figure~\ref{Figure_2D_All_Module_Effect} explores the choice of JL dimension, $p$, for the datasets in Study I. The effects of $p$ on both the computation cost per iteration and the SER benefit per iteration combine to yield very similar performance across the range $0.5 N_c \leq p \leq 2N_c$. Note that even when $p=1$, HICU still manages to reconstruct. This implies that randomness through JL projection can efficiently capture most of the information in the large nullspace. The bottom panel in Figure~\ref{Figure_2D_All_Module_Effect} suggests that the impact of the CO strategy slightly exceeds the impact of JL on computing time, and that the two speedup factors are approximately multiplicative when the two numerical strategies are adopted jointly.

\begin{figure}[!ht]
    \centering
    \includegraphics[width = 16.5cm]{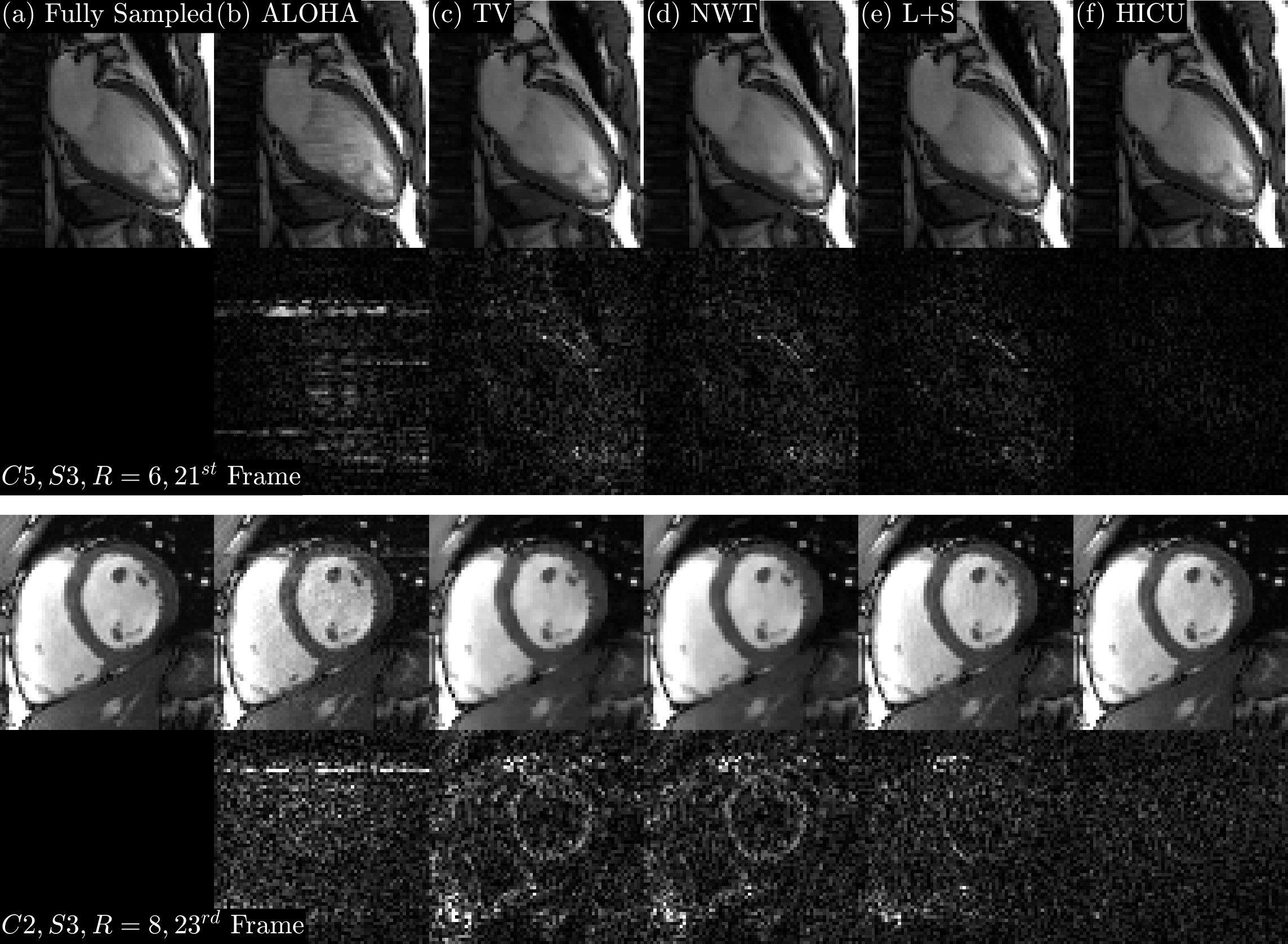}
    \caption{The reconstruction results of all methods for the $21^{\text{st}}$ frame of ($C5, S3, R=6$) and $23^{\text{rd}}$ frame of ($C2, S3, R=8$).  First and third rows: SSoS image. Second and fourth rows: $5.34\,\times$\,absolute error.}
    \label{Figure_2D_T_Images}
\end{figure}

For 2D+t images in Study II, Figures~\ref{Figure_nD_SER} and \ref{Figure_2D_T_Images} show that HICU can provide consistently better SER than SENSE-based methods. The SER gain may be attributed to a low-rank model better capturing the multi-coil structure than do the sensitivity maps extracted from the time-averaged ACS region. The average computing times in Study II of the fastest SENSE-based method, NWT, and HICU are $78.5$\,s and $350.5$\,s. Yet, to achieve only the same average SER of the best SENSE-based method, L+S, HICU can compute in $77.7$\,s using only $I=16$ iterations. ALOHA reconstructions averaged $4.1$\,dB less SER and required over $13$ times longer computation than HICU. The ALOHA reconstruction in Figure~\ref{Figure_2D_T_Images} shows banding artifacts, whereas others do not; this difference may be attributed to the slice by slice processing in ALOHA.

For 3D knee images in Study III, Figures~\ref{Figure_nD_SER} and \ref{Figure_3D_Images} show that HICU can provide, on average, $0.68$\,dB improvement in SER and $23\%$ less computation time than the SENSE-based methods reported.
Thus, in both Study II and the volumetric imaging application in Study III, HICU matrix completion enables calibrationless imaging to exceed SENSE-based CS imaging in both SER and computing speed.  In Study IV, HICU is demonstrated for multi-shot diffusion weighted imaging on a large k-space array of $74.7$\,million samples.  Figure~\ref{Figure_MSDWI_Image} demonstrates qualitatively similar results to IRLS MUSSELS, with five times shorter computation time.

The optimization task addressed by HICU is found in LORAKS \cite{haldar2014} and is similar to many other calibrationless parallel imaging approaches.  Consequently, the numerical techniques (rSVD, JL, CO) employed in HICU could likewise accelerate algorithms used to optimize other variations of cost functions. For example, for the case of ($S1$, $R=3$) in Study I, the use of rSVD (with projection dimension $3r$) reduces average computation time, T$_c$, for SAKE from $239.1 \pm 71.7$\,s to $111.9 \pm 32.7$\,s, without any performance loss. Moreover, various forms of regularization may be used as the functions $f_i(\bm{x})$ found in [\ref{eq:hicu_cost2}]  but are not considered in the results presented here. For example, in addition to a soft data consistency function, the $\ell_1$ norm of a linear sparsifying transform of $\mathbb{X}$ results in a simple soft-thresholding after Step 8 of Algorithm \ref{alg:h2_new}. Similarly, a learned plug-in denoiser could be employed \cite{AhmadPNP, zhao2020calibrationless}. In either case, this additional regularizer operates on the premise that multi-coil image structure not captured by the low-rank linear dependence assumption can be exploited via a nonlinear denoising step. For the least-squares subproblem, gradient descent with ELS is observed faster than LSQR and conjugate gradient descent in our sparse and rectangular linear system.  

Although they attempt to optimize different cost functions, HICU and GIRAF algorithmically share much in common.  Instead of the cost in [\ref{eq:l2tail}], GIRAF seeks to minimize the smoothed Schatten-$q$ quasi-norm,
\begin{equation}
    \label{eq:giraf_cost}
    J_{\text{G}} ( \bm{x} )  = \sum_{k =1}^{n}
    \left( \sigma^2_k \left\{ \mathcal{H}_{\mathbb{K}} \left(  \mathbb{X}  \right)  \right\} + \epsilon \right)^{q/2}.
\end{equation}
In its iterative reweighted least-squares computation \cite{Sullivan1987,Fornasier2011,mohan}, GIRAF uses weighted right singular vectors of $\bm{H}(\bm{x})$ as annihilating kernels; the $k^{\text{th}}$ singular vector is weighted by $w_k = (\sigma_k + \epsilon )^{-(1-q/2)/2}$, $0 \leq q \leq 1$. Seen in this framework, HICU uses weights $w_k=0$ for $k\leq r $ and $w_k=1$ for $r+1 \leq k \leq n$. GIRAF requires computation of the full set of $m$ singular values at each iteration, whereas HICU only requires 
the $r$ principal right singular vectors, from which a null space basis is found via Householder reflections. Therefore, HICU can benefit from rSVD versus SVD of the Gram matrix. For Study I, we did not observe appreciable difference in computing time between rSVD and SVD of Gram matrix. However, rSVD was significantly faster than SVD of the Gram matrix for larger applications considered in Studies II, III, and IV. For example, for Study II, when rSVD was replaced with SVD of Gram matrix, the average computation time increased by $52\%$.

The current implementation of HICU has several limitations.  HICU requires, like related Cl-MRI algorithms, several tuning parameters, including: rank, kernel support, size of the central region in CO, and the number of iterations or, equivalently, a stopping criterion. An automated selection of these parameters, especially $r$, is a direction for future work.

\section{CONCLUSION}
A variety of physical features contribute to the approximate linear dependence among neighboring k-space samples. This dependence has been leveraged to yield many existing algorithms to recover unobserved k-space samples. We build on this rich literature to present a computational approach, HICU, that is simple, fast, memory efficient, and directly extensible to imaging in higher dimensions. For structured low-rank matrix completion, HICU iteratively estimates unknown annihilating kernels and k-space samples to minimize the tail sum of squared singular values.  Computational speed is achieved by random projection, at each iteration, of the annihilating kernel space to a lower-dimensional subspace and employing a small, high-SNR center of k-space to quickly build initial estimates of nullspace vectors.  Results from 2D brain imaging show that HICU can offer over an order of magnitude speedup compared to existing algorithms.  Results from 2D+t and 3D imaging show that HICU can make calibrationless imaging computationally comparable to SENSE-based CS methods while providing improved reconstruction SER. Results from 2D, 2D+t, 3D, and MSDWI demonstrate that HICU can generalize to various MRI applications.

\section{ACKNOWLEDGMENTS}
This work was funded in part by NIH project R01HL135489. The authors are grateful to Merry Mani and Mathews Jacob for providing the DWI dataset and code used in Study IV and thank Kaiying Xie and Chong Chen for valuable discussions. The authors are also thankful to Martin Uecker and Christian Holme for assistance in optimizing parameters in the NLINV code. In memory of FEMD, 1927-2020.
\clearpage

\bibliography{HICU.bib}
\clearpage

\section{FIGURES CAPTIONS}

\clearpage

\clearpage

\clearpage

\clearpage

\clearpage

\clearpage

\clearpage

\clearpage

\section{TABLES}
\begin{algorithm}[!ht]
\caption{HICU Multi-level Hankel Matrix Completion}
\label{alg:h2_new}
\begin{algorithmic}[1]
    \REQUIRE 
    Zero-filled observed k-space, $\mathbb{X}_0$;
    Sampling mask, $\mathbb{M}$;
    Kernel mask, $\mathbb{K}$; 
    Region of k-space, $\mathbb{S}^{(i)}$;
    Rank, $r$; 
    Compression dimension, $p$;
    Number of iterations, $I$; 
    Number of gradient steps per iteration, $G^{(i)}$;
    Initialize $\bm{x}^{(0)} = \text{vec} \{ \mathbb{X}_0 \}$
    \FOR{$i = 1 \text{ to } I$}
    \STATE Compute $\bm{V}^{(i)}, r \text{ principal right singular vectors of } \bm{H}^{(i-1)} \left( \bm{x}^{(i-1)} \right)$ via rSVD \hfill{{\bf rSVD}}\\
    \STATE Compute orthonormal basis, $\bm{Q}^{(i)}$, orthogonal to $\bm{V}^{(i)}$ via $r$ Householder reflections
    \STATE Select region, $\mathbb{S}^{(i)}$, on which to compute valid convolution using $\bm{H}^{(i)} \left( \bm{x}^{(i-1)} \right)$ \hfill{{\bf CO}}\\
    \STATE Prepare for $G^{(i)}$ descent steps, $\bm{y}^{(0)} = \bm{x}^{(i-1)}$
    \FOR{$j = 1$ \text{to} $G^{(i)}$}
    \STATE Compress nullspace to $p$ dimensions, $\widetilde{\bm{Q}}^{(i,j)} =  \bm{Q}^{(i)} \bm{P}^{(i,j)}$, where $\bm{P}^{(i,j)}$ is i.i.d.\ normal \hfill{{\bf JL}}\\
    \STATE Compute gradient,
    $\bm{g}^{(j)} =  \sum_{k=1}^{p}  \nabla_{\bm{y}} \left\| 
    \bm{H}^{(i)} ( \bm{y}^{(j-1)} )  \widetilde{\bm{q}}_{k}^{(i,j)} \right\|^2 + \sum_{i=1}^{c} \lambda_c \nabla_{\bm{y}} f_i ( \bm{y}^{(j-1)} )$ 
    \STATE Update with step size set by exact line search,
    $\bm{y}^{(j)} = \bm{y}^{(j-1)} - \eta^{(j)} \bm{g}^{(j)}$ \hfill{{\bf ELS}}\\
    \ENDFOR\\
    \STATE Save result of gradient steps, $\bm{x}^{(i)} = \bm{y}^{\left(G^{(i)}\right)}$
    \ENDFOR\\
    \ENSURE Reconstructed k-space, $\bm{x}^{(I)}$
\end{algorithmic}
\end{algorithm}
\clearpage

\begin{table}[!ht]
\centering
\scriptsize
\setlength\tabcolsep{2.7pt} 
\begin{tabular}{|c|c|c|c|c|c|c|c|c|}
\hline
$2D$    & \multicolumn{4}{c|}{$S1,R=3$}                                                                             & \multicolumn{4}{c|}{$S1,R=5$}                                                                             \\
\hline
        & SER (dB)                  & SSIM                      & HFEN (dB)                 & T$_c$ (s)             & SER (dB)                  & SSIM                      & HFEN (dB)                 & T$_c$ (s)             \\
\hline
NLINV   & 12.62$\pm$0.53            & 0.575$\pm$0.029           &  9.82$\pm$0.49            &  18.2$\pm$1.7         & 5.20$\pm$0.44             & 0.422$\pm$0.028           & -3.69$\pm$0.57            & 21.7$\pm$1.3          \\
\hline
SAKE    & 17.62$\pm$0.63            & \textbf{0.715$\pm$0.030}  & \textbf{14.13$\pm$0.84}   & 239.1$\pm$71.7        & \textbf{14.73$\pm$0.59}   & 0.634$\pm$0.036           & \textbf{-11.02$\pm$0.73}  & 966.2$\pm$486.6       \\
\hline
LORAKS  & \textbf{17.73$\pm$0.63}   & \textbf{0.715$\pm$0.030}  & \textbf{14.13$\pm$0.84}   &  84.9$\pm$6.6         & 14.69$\pm$0.60            & 0.634$\pm$0.037           & -10.99$\pm$0.72           & 566.7$\pm$156.6       \\
\hline
HICU    & 17.50$\pm$0.60            & 0.714$\pm$0.030           & 13.87$\pm$0.83            & \textbf{4.4$\pm$0.5}  & 14.59$\pm$0.56            & \textbf{0.640$\pm$0.036}  & -10.92$\pm$0.72           & \textbf{13.8$\pm$2.9} \\
\hline
\hline
$2D+t$  & \multicolumn{4}{c|}{$S2,R=6$}                                                                             & \multicolumn{4}{c|}{$S2,R=8$}                                                                             \\
\hline
        & SER (dB)                  & SSIM                      & HFEN (dB)                 & Time (s)              & SER (dB)                  & SSIM                      & HFEN (dB)                 & Time (s)              \\
\hline        
ALOHA   & 24.80$\pm$1.37            & 0.958$\pm$0.017           & 24.70$\pm$0.68            & 4830$\pm$495          & 21.94$\pm$1.88            & 0.948$\pm$0.013           & -21.00$\pm$3.17           & 4532$\pm$446          \\
\hline
TV      & 25.45$\pm$1.10            & 0.959$\pm$0.013           & 25.29$\pm$1.29            & 87.6$\pm$8.2          & 24.17$\pm$0.81            & 0.957$\pm$0.011           & -23.73$\pm$1.48           & 87.6$\pm$9.1          \\
\hline
NWT     & 25.90$\pm$1.20            & 0.956$\pm$0.013           & 25.64$\pm$1.31            & \textbf{78.2$\pm$7.4} & 24.78$\pm$0.96            & 0.955$\pm$0.012           & -24.20$\pm$1.50           & \textbf{78.7$\pm$8.4} \\
\hline
L+S     & 26.16$\pm$1.27            & 0.960$\pm$0.013           & 25.86$\pm$1.45            & 90.1$\pm$13.3         & 25.15$\pm$1.08            & \textbf{0.959$\pm$0.011}  & -24.59$\pm$1.58           & 90.2$\pm$13.3         \\
\hline
HICU    & \textbf{27.90$\pm$1.58}   & \textbf{0.964$\pm$0.013}  & \textbf{27.29$\pm$1.10}   &350.5$\pm$31.5         & \textbf{27.19$\pm$1.47}   & \textbf{0.959$\pm$0.014}  & \textbf{-26.48$\pm$1.05}  &350.4$\pm$30.8         \\
\hline
\hline
$3D$    & \multicolumn{4}{c|}{$S3,R=4.86$}                                                                          & \multicolumn{4}{c|}{$S4,R=5$}                                                                             \\
\hline
        & SER (dB)                  & SSIM                      & HFEN (dB)                 & Time (s)              & SER (dB)                  & SSIM                      & HFEN (dB)                 & Time (s)              \\
\hline        
NLINV   & 10.38$\pm$2.22            & 0.619$\pm$0.120           & 6.87$\pm$1.48             & 657.4$\pm$14.3        & 9.81$\pm$1.95             & 0.698$\pm$0.113           & -6.67$\pm$1.35            & 653.1$\pm$11.7        \\
\hline
WT      & 10.68$\pm$2.33            & 0.723$\pm$0.105           & 7.06$\pm$1.46             & 695.3$\pm$23.2        & N/A                       & N/A                       & N/A                       & N/A                   \\
\hline
HICU    & \textbf{11.05$\pm$2.17}   & \textbf{0.826$\pm$0.069}  & \textbf{7.37$\pm$1.25}    & \textbf{515.1$\pm$2.2}& \textbf{10.68$\pm$2.11}   & \textbf{0.822$\pm$0.075}  & -\textbf{7.12$\pm$1.25}   & \textbf{515.7$\pm$1.6}\\
\hline
\hline
MSDWI   & \multicolumn{8}{c|}{$S5,R=6.74$}                                                                                                                                                                                      \\ 
\hline
        & \multicolumn{2}{c|}{SER (dB)}                         & \multicolumn{2}{c|}{SSIM}                         & \multicolumn{2}{c|}{HFEN (dB)}                        & \multicolumn{2}{c|}{Time (s)}                     \\ 
\hline
MUSSELS & \multicolumn{2}{c|}{N/A}                              & \multicolumn{2}{c|}{N/A}                          & \multicolumn{2}{c|}{N/A}                              & \multicolumn{2}{c|}{5397.9 }                      \\ 
\hline
HICU    & \multicolumn{2}{c|}{N/A}                              & \multicolumn{2}{c|}{N/A}                          & \multicolumn{2}{c|}{N/A}                              & \multicolumn{2}{c|}{\textbf{1061.9}}              \\ 
\hline
\end{tabular}
\caption{2D: For sampling pattern ($S1,R=3,5$), k-space SER and SSIM, HFEN of reconstruction of SAKE, LORAKS, HICU at one hour, and NLINV at the final iteration. The computing time T$_c$ to reach within $0.1$ dB less than the one hour results 
(computing time fro NLINV). 2D+t: For sampling pattern ($S2,R=6,8$), k-space SER and SSIM, HFEN, and computing time for ALOHA, TV, NWT, L+S, and HICU. 3D: For sampling patterns ($S3,R=4.86, S4,R=5$), k-space SER and SSIM, HFEN, and computing time for NLINV, WT, and HICU.}
\label{Tab:nD SER SSIM HFEN}
\end{table}
\scriptsize
\clearpage

\section{SUPPORTING INFORMATION}

\renewcommand{\thefigure}{\arabic{figure}}
\captionsetup[figure]{name=Supporting Information Video S\!\!}
\setcounter{figure}{0}

\begin{figure}[h]
	\centering
	\caption{The reconstructed cardiac cine for ($C5, S3, R=6$) along with the fully sampled reference. First and third rows: SSoS image series. Second rows: $5.3\,\times$\,absolute error.}
	\label{video 1}
\end{figure}

\begin{figure}[h]
	\centering
	\caption{The reconstructed cardiac cine for ($C2, S3, R=8$) along with the fully sampled reference. First and third rows: SSoS image series. Second rows: $5.3\,\times$\,absolute error.}
	\label{video 2}
\end{figure}



\end{document}